# Accessing Different Spin-Disordered States using First Order Reversal Curves

Randy K. Dumas,[1,2,*] Peter K. Greene,[1] Dustin A. Gilbert,[1] Li Ye,[1] Chaolin Zha,[3]

Johan Åkerman,[2,3] and Kai Liu[1,*]

[1]Department of Physics, University of California, Davis 95616, USA

[2]Department of Physics, University of Gothenburg, Gothenburg 412 96, Sweden

[3]Materials Physics, School of ICT, Royal Institute of Technology (KTH), Kista 164 40, Sweden

**Abstract**

Combined first order reversal curve (FORC) analyses of the magnetization (M-FORC) and magnetoresistance (MR-FORC) have been employed to provide a comprehensive study of the M-MR correlation in two canonical systems: a NiFe/Cu/FePt pseudo spin-valve (PSV) and a [Co/Cu]$_8$ multilayer.  In the PSV, due to the large difference in switching fields and minimal interactions between the NiFe and FePt layers, the M and MR show a simple one-to-one relationship during reversal.  In the [Co/Cu]$_8$ multilayer, the correlation between the magnetization reversal and MR evolution is more complex.  This is primarily due to the similar switching fields of, and interactions between, the constituent Co layers.  The FORC protocol accesses states with much higher spin disorders and larger MR than those found along the conventional major loop field-cycle.  Unlike the M-FORC measurements, which only probe changes in the macroscopic magnetization, the MR-FORCs are more sensitive to the microscopic domain configurations, as those are most important in determining the resultant MR effect size. This approach is generally applicable to spintronic systems to realize the maximum spin-disorder and the largest MR.

**PACS's: 75.70.-i, 75.47.De, 73.50.Jt, 75.60.Jk**

# I. INTRODUCTION

Uncovering the mechanisms that govern hysteretic reversal, particularly in magnetically heterogeneous systems, is critical to their basic understanding and potential applications. Of particular interest are ferromagnetic/non-magnetic (FM/NM) layered structures exhibiting either the giant magnetoresistance (GMR)[1,2] or the tunneling magnetoresistance (TMR)[3] effect, which has enabled a host of exciting spintronic devices, such as hard disk drive read heads and magnetic random access memory (MRAM).[4-6] It is well known that the magnitude of the magnetoresistance (MR) effect is intimately dependent on the spacer layer thickness and the interlayer exchange coupling. For example, the MR of Co/Cu multilayers has been found to oscillate[7] with a period of ~1 nm as the thickness of the NM spacer layer is increased due to an oscillatory exchange coupling mechanism.[8] Therefore, by tuning the thickness of the NM layer to promote a preferential anti-parallel coupling, the MR can be maximized. For a given spintronic system, the achievable MR magnitude critically depends on the magnetization configurations of the constituent FM entities, where a low (high) resistance state is realized for a spin ordered (disordered) state. For example, in GMR systems, be it multilayer thin films or granular solids, the MR scales with $<cos\ \phi_{ij}> \sim <cos\ \theta>^2 \sim (M/M_S)^2$, where $\phi_{ij}$ is the angle between the magnetizations of the FM entities (an indicator of the spin disorder), and $\theta$ is the angle between the magnetization direction of a FM entity and the applied field.[9] A variety of techniques have been employed to study the microscopic domain configurations in heterogeneous multilayers, correlating with other magnetic properties, including e.g., polarized neutron reflectometry, scanning electron microscopy with polarization analysis, and element-specific x-ray magnetic dichroism.[10-12] Prior experiments have revealed that the virgin state MR of an as-prepared sample can actually be significantly larger than that accessible after subsequent field cycling during a major loop measurement.[10, 13] These experiments have demonstrated that while there is a high degree of *correlated* anti-parallel alignment across the NM spacer in the as-prepared state, during a major loop measurement the degree of domain correlation becomes more random at the coercive field[10, 13] leading to a reduction in the maximum achievable MR. Furthermore, other coupling mechanisms such as exchange coupling via pinholes,[14] Néel (orange peel) coupling,[15] and magnetostatically driven domain replication[16] are often present in spin valves and may also play a significant role. More specifically, these results highlight the critical role that specific microscopic magnetic domain configurations have on the resultant MR. Given these complexities, precisely how the magnetization and MR will correlate during reversal is not always straightforward.

In addition to providing a useful qualitative "fingerprint"[17-20] of the reversal mechanism, first order reversal curve (FORC)[21, 22] analysis has shown the ability to probe a wealth of quantitative information regarding reversible/irreversible switching,[23-28] interactions,[29-34] and distributions[35, 36] of key magnetic characteristics not readily accessible from standard major loop or remanence curve investigations. The FORC measurement involves taking numerous partial (minor) hysteresis loops, changing the energy landscape in such a way so as to allow the system to reconfigure along paths different from the conventional major loop.[37] For example, in perpendicular magnetic anisotropy systems, it has been found that certain FORCs actually protrude outside of the major loop.[38, 39] While the FORC technique has been primarily utilized to study magnetic hysteresis, its applicability has been extended to systems exhibiting thermal,[40] electrochemical,[41] ferroelectric,[42] and resistive[43-45] hysteresis as well.

In this work, we have carried out a *combined* FORC analysis of the magnetization, termed M-FORC, *and* magnetoresistance, termed MR-FORC, to provide a comprehensive picture of the reversal mechanisms in two canonical magnetoresistive systems: a pseudo spin-valve (PSV) and a classic multilayered Co/Cu GMR film stack. We find that the FORC methodology allows us to access different spin disordered states, even achieving MR values much larger than the maximum along the conventional magnetic field cycle.

## II. EXPERIMENT

Samples for this study were deposited on Si substrates using magnetron sputtering in a vacuum chamber with a base pressure of better than $1\times10^{-8}$ Torr and at an Ar working pressure of 5 mTorr. The PSV has the following layer structure (thicknesses in nm): Ta(6)/Pt(3)/$Fe_{53}Pt_{47}$(20)/$Co_{50}Fe_{50}$(1.5)/Cu(4.5)/$Co_{50}Fe_{50}$(2)/$Ni_{81}Fe_{19}$(3)/Ta(5). The $Fe_{53}Pt_{47}$, referred to as FePt hereafter, was deposited by co-sputtering high purity Fe and Pt targets at a substrate temperature of 700 ºC and then annealed *in situ* for 10 minutes, to promote formation of the high anisotropy $L1_0$ phase. The film was then allowed to cool to room temperature in vacuum before deposition of the subsequent layers, in order to minimize interdiffusion. The $Co_{50}Fe_{50}$ and $Ni_{81}Fe_{19}$, henceforth referred to simply as CoFe and NiFe, were deposited from stoichiometric targets. The thin spin polarizing CoFe insertion layers are introduced to increase the MR ratio and rigidly reverse with the neighboring FePt or NiFe. For the rest of the paper we will use FePt and NiFe to refer to the CoFe/FePt and CoFe/NiFe bilayer, respectively. Additional structural characterizations, e.g. x-ray diffraction, and atomic force microscopy, can be found in Refs. 46, 47. The polycrystalline GMR multilayer stack, [Co(1nm)/Cu(3nm)]$_8$, was modeled after the samples presented in Ref. 7. The Cu spacer layer thickness was tuned to the third MR oscillation peak,[7] which leads to antiferromagnetic interlayer exchange coupling of the Co layers and a still sizeable MR.

Magnetization measurements were performed at room temperature using a Princeton Measurements Corp. 2900/3900 alternating gradient and vibrating sample magnetometer (AGM/VSM) with the applied field in the plane of the films. The current-in-plane (CIP) MR was measured in the transverse geometry using standard 4-probe techniques with the applied field in the plane of the films and perpendicular to the current flow. The MR at a given field, $H$, is expressed as a percentage relative to the resistance, $R$, in a saturating field, $H_s$, as follows: $MR(H)=[R(H)-R(H_s)]/R(H_s)\times 100\%$.

FORCs are measured by the following procedure. After positive saturation the applied field is reduced to a given reversal field, $H_R$. From this reversal field the magnetization or resistance is then measured as the applied field, $H$, is swept back towards positive saturation, thereby tracing out a single FORC. This process is repeated for a series of decreasing reversal fields creating a family of FORCs. The FORC distribution is then defined as a mixed second order derivative of the family of FORCs:

$$\rho(H, H_R) \equiv -\frac{1}{2}\frac{\partial^2 \alpha(H, H_R)}{\partial H \partial H_R}, \qquad (1)$$

where $\alpha$ corresponds to either the normalized magnetization, $M/M_S$, or the magnetoresistance, *MR*. The multiplicative factor of 1/2 is typically included in M-FORC analysis for normalization purposes and the *negative* sign reflects the fact that the FORCs are measured from $H_R$ values originating on the *descending* branch of the major loop. However, since direct comparisons between M-FORC and MR-FORC distributions will be drawn, we will normalize by their respective maximum values and the multiplicative factor will not be critical to the discussions presented here. The M-FORC and MR-FORC distributions are plotted against ($H$, $H_R$) coordinates on a contour map where $H \geq H_R$ by design. Following the measurement procedure the FORC distribution is read in a "top-down" fashion and from left to right for a particular reversal field. The FORC distribution provides a useful "fingerprint" of the reversal mechanism by mapping out, in ($H$, $H_R$) coordinates, only the *irreversible* switching processes. It is often useful to evaluate the irreversible switching using the FORC-switching field distribution (FORC-SFD). This is accomplished by projecting the FORC distribution onto the $H_R$-axis, equivalent to an integration over the applied field $H$:[25, 48]

$$\int \frac{\partial^2 \alpha(H, H_R)}{\partial H \partial H_R} dH = \frac{d\alpha(H_R)}{dH_R} \qquad (2)$$

## III. RESULTS

### A. Major Loop Analysis

We begin with a standard major loop analysis. The magnetization (solid squares) and MR (open circles) major hysteresis loops for the FePt/Cu/NiFe PSV and [Co/Cu]$_8$ multilayer are shown in Figs. 1(a) and 1(b), respectively. The magnetization loop for the FePt/Cu/NiFe PSV, Fig. 1(a), shows two clear and separate switching events corresponding to the reversal of the soft NiFe ($H= \pm 0.3$ kOe) and hard FePt ($H= \pm 4.7$ kOe) bi-layers. The MR curve, referred to as the MR major loop, exhibits two corresponding plateaus and a clear MR maximum of 4.1%, due to the anti-parallel alignment of the NiFe and FePt.

The magnetization reversal behavior of the [Co/Cu]$_8$ multilayer, Fig. 1(b), is qualitatively different from the PSV, showing only a single dominant switching event. The MR reaches a peak with a maximum of 4.2%. As is commonly observed in such systems, the maximum MR when the magnetic field is cycled along the major loop does *not* correspond to a fully demagnetized state (i.e. the zero magnetization at the coercive fields $\pm H_C$), where one might expect the spin disorder to be at a maximum as well. In fact, the MR at $\pm H_C$ is only 4.0%, lower than the MR major loop maximum of 4.2%. Previously, polarized neutron reflectivity measurements have revealed that an uncorrelated domain structure in similar Co/Cu multilayers leads to a reduced MR at the coercive fields.[10]

**B. FORC Analysis**

The family of M-FORCs and the corresponding M-FORC distribution, calculated using Eqn. (1), for the FePt/Cu/NiFe PSV are shown in Figs. 2(a) and 2(b), respectively. The M-FORC distribution, Fig. 2(b), is characterized by two primary features. The first, highlighted with a dashed circle, is a very sharp and highly localized positive peak that occurs for slightly negative $H_R$ values. This feature is caused by the irreversibility associated with the initial rapid switching of the soft NiFe. The second feature, highlighted with an oval, occurs for more negative $H_R$ values and can be identified with the more gradual irreversible switching of the hard FePt. The family of MR-FORCs and corresponding MR-FORC distribution are shown in Figs. 2(c) and 2(d), respectively, where the latter now shows three primary features. The first, highlighted with a dashed circle, is a highly localized *negative* peak associated with the rapid increase in MR for small negative $H_R$ values, as the NiFe switches anti-parallel to the FePt. Unlike what was observed in the M-FORC distribution, this peak in the MR-FORC distribution is negative because, for -0.5 <$H_R$<0 kOe, the MR-FORC slope *decreases* at successively more negative $H_R$ in the field range of 0 < $H$ < 0.5 kOe [Fig. 2(c) inset]. The location of the second feature (highlighted with a black rectangle) corresponds to the irreversible switching of the hard FePt layer *and* the soft NiFe layer. For reversal fields where the FePt has partially switched (-6 kOe < $H_R$ <-4 kOe), as the applied field is increased the resistance first increases as the FePt magnetization becomes more anti-parallel to the NiFe. As the applied field continues to increase the resistance now shows a rapid irreversible decrease as the NiFe switches back towards positive saturation for small positive fields, thus defining the location of this boxed feature along the $H$-axis. It is interesting to note that while this feature is extended along the $H_R$-axis, due to the broad switching field distribution of the FePt, it exhibits a relatively narrow extent along the $H$-axis due to the rapid switching of the NiFe. An example FORC for $H_R$=-4.8 kOe can be seen in Fig. 4(a). Finally, a third MR-FORC featured highlighted with a black oval corresponds to the switching of the FePt alone, and closely mimics the shape of the corresponding M-FORC feature shown in Fig. 2(b). Note that this feature is now negative as the slope progressively increases for more negative $H_R$ in the field range 1 kOe < $H$ < 6 kOe. In order to provide a more clear comparison between the M-FORC and MR-FORC distributions, the normalized FORC-SFDs, calculated using Eqn. (2), are shown in Fig. 2(e). Each FORC-SFD clearly exhibits two distinct peaks that can be separately associated with the independent switching of the NiFe (centered at $H_R$= -0.3kOe) and FePt (centered at $H_R$= -4.7 kOe) bi-layers. More importantly, however, is that the irreversibility in the magnetization and MR track each other "in-phase" as $H_R$ is decreased. Other than a sign difference and relative amplitudes of the FORC-SFD peaks, there is a simple and direct one-to-one correspondence between irreversibility exhibited in the magnetization and MR response.

The behavior of the [Co/Cu]$_8$ multilayer is markedly different. The family of M-FORCs and corresponding M-FORC distribution for the [Co/Cu]$_8$ GMR stack are shown in Figs. 3(a) and 3(b), respectively. Interpretation of the M-FORC distribution is less straightforward here because, unlike for the FePt/NiFe PSV, features cannot be easily linked to a given magnetic layer, but instead manifest the irreversible switching processes of the film as a whole. The family of MR-FORCs and the corresponding MR-FORC distribution are shown in Figs. 3(c) and 3(d), respectively. Interestingly, the MR-FORC features, Fig. 3(d), show peaks and valleys in very different locations as compared to the M-FORC, Fig. 3(c), indicative that a drastically different irreversibility landscape is extracted from the MR-FORCs, as further verified in the FORC-SFDs, shown in Fig. 3(e). The MR-FORC SFD exhibits a negative as well as a positive peak, similar to that in the aforementioned FePt/Cu/NiFe PSV case, whereas the M-FORC SFD shows only a single peak coincident with the single switching event exhibited by the major loop. The FORC-SFDs shown in Fig. 3(e) are now highly "out-of-phase", demonstrating that the irreversibility in the magnetization and MR no longer show a simple correlation. Finally, another striking behavior of [Co/Cu]$_8$ GMR sample is that along selected MR-FORCs, Fig. 3(c), MR values up to 4.8% are found, larger than the 4.2% maximum of the MR major loop.

The most interesting differences between the reversal behavior of the FePt/Cu/NiFe PSV and the [Co/Cu]$_8$ multilayer are best highlighted by considering their MR-FORCs. In order to better visualize the reversal, only three selected MR-FORCs, as well as the MR major loop, for the FePt/Cu/NiFe PSV and the [Co/Cu]$_8$ multilayer are shown in Figs. 4(a) and 4(b), respectively. The three MR-FORCs correspond to $H_R$ values just before, on, and after the major loop MR maximum [points 1-3 on Figs. 4(a) and 4(b), respectively]. In the PSV case, for the MR-FORCs starting at $H_R$= -0.3 and -1.0 kOe, it is only the NiFe layer that has undergone reversal. Therefore the MR simply decreases as the NiFe approaches positive saturation and parallel alignment with the FePt is restored. Note that for the FORC at $H_R$= -1.0 kOe, it starts with the maximum spin disorder [point 2 of Fig. 4(a)] as the NiFe is completely opposite to the FePt; thus the MR decreases from a global maximum of 4.1%. For the MR-FORC starting at $H_R$=-4.8 kOe [point 3 of Fig. 4(a)], the NiFe layer has undergone a complete reversal while the FePt has only partially reversed. As the applied field $H$ is increased, the MR will first increase as a greater fraction of the FePt again becomes predominantly anti-parallel to the NiFe. However, the degree of spin disorder never exceeds that at point 2, and the MR value along that FORC is well below the global maximum of 4.1%. Once the applied field reaches the switching field of the NiFe, the MR first drops precipitously as the NiFe quickly aligns in the +$H$ direction ($0 < H <0.5$ kOe) and then slowly returns to zero as the FePt continues its return to positive saturation ($H > 0.5$ kOe). The reversal process is simple because the NiFe and FePt switch in very different fields and do not significantly interact. Therefore along the major loop, a perfectly anti-parallel configuration of the NiFe and FePt is possible, and hence the maximum achievable MR. This also manifests itself in the observed MR-FORCs, which all lie within the bounds of the major loop, and in the FORC-SFDs, Fig. 2(e), which show a simple relation between the irreversibility in the magnetization and MR.

The behavior of the [Co/Cu]$_8$ multilayer is much more complex. Most notably, a subset of the measured MR-FORCs actually protrudes outside the MR major loop. Furthermore, the maximum MR observed during the MR-FORC measurements, 4.8%, does not originate from the peak of the MR major loop, i.e., from $H_R$= -64 Oe [point 2 in Fig. 4(b)], but from a significantly more negative value of $H_R$= -143 Oe [point 3 in Fig. 4(b)].

## IV. SIMULATIONS AND DISCUSSIONS

To gain further insights into the magnetization reversal behavior in the Co/Cu system, micromagnetic simulations have been performed using the 3D Oxsii OOMMF simulation platform.[49] The simulations modeled a simple tri-layer system with lateral dimensions of 2×2 µm$^2$ and a vertical structure of Co(2 nm)/spacer(2 nm)/Co(2 nm), discretized into $4 \times 4 \times 2$ nm$^3$ tetragonal cells. The large lateral dimensions were chosen to allow for multi-domain reversal.

Material parameters suitable for polycrystalline bulk Co were used (saturation magnetization $M_S=1.4\times10^6$ A/m, exchange stiffness $A=3.0\times10^{-11}$ J/m) and crystalline anisotropy was neglected. The non-magnetic spacer was considered magnetically inert. Furthermore, a randomly distributed crystalline defect density (0.1 %) was included to promote domain nucleation. Each defect site shares the same material properties as the surrounding Co layer except that the defect cells have a (110) cubic magnetocrystalline easy axis with an anisotropy $K=0.52$ MJ/m$^3$ corresponding to the uniaxial anisotropy of bulk Co. A long range bilinear exchange interaction is included in the simulations with a surface exchange coefficient of $\sigma = -8.0 \times 10^{-6}$ J/m$^2$ to simulate antiferromagnetic exchange coupling across the spacer.[50] Finally, the total normalized cross-spacer spin correlation is evaluated pair-wise and cell-by-cell as the dot product of unit moments $\hat{m}$ which lie directly across from one another in the top and bottom Co layers, $S = \frac{\sum \hat{m}_i^{top} \cdot \hat{m}_i^{bottom}}{total\ number\ of\ pairs}$ (within ±1). The resultant total cross-spacer spin disorder, defined as $D=(1-S)/2$ ($D=1$ and 0 for complete disordered and ordered state, respectively), then provides a relative measure of the MR.[51]

A simulated major loop and three representative MR-FORCs for the Co/Cu/Co tri-layer structure are shown in Fig. 4(c) and qualitatively reproduce the experimental results shown in Fig. 4(b).[52] The simulated magnetization profiles in the top and bottom Co layers are shown in the upper right inset of Fig. 4(c) along the major loop for an applied field of $H$=-23 Oe, indicated with a black dot, which has the maximum spin disorder along the MR major loop. While there is a large degree of anti-parallel alignment of the Co magnetizations across the spacer, over a sizable region the moments remain partially aligned, particularly the upper and lower edges of the simulated sample, limiting the maximum spin disorder realized to $D$~0.8. In contrast, the MR-FORC originating from a reversal field of $H_R$=-46 Oe is able to access a state where the spin disorder is much larger than that found along the major loop. Nearly ideal spin disorder ($D \sim 1$) is achieved along this particular FORC as a different path towards positive saturation is taken. This is highlighted by comparing the M and MR-FORCs as well as the domain images as shown in Fig. 5(a) main panel and inset, respectively. At $H=H_R=-46$ Oe the M-FORC is near a negatively saturated state with the magnetizations of both top and bottom layers pointing primarily to the left (defined as the negative field direction). As $H$ is increased the moments of both top and bottom layers become nearly perfectly anti-parallel at $H=-3$ Oe, exhibiting a zero remanence state ($M/M_S=0$) and a corresponding maximum in the total spin disorder. Interestingly, the experimental M and MR-FORCs for the Co/Cu multilayer, Fig. 5(b), show a qualitatively similar behavior. Namely, reversing at $H_R=-143$ Oe, the M-FORC starts out from a more negatively saturated state, and along the MR-FORC the maximum occurs at $H=40$ Oe, corresponding to a completely demagnetized state. Note that at the coercive field of the major loop, where the sample is also in a demagnetized state, the MR is only 4.0%, Fig. 1(b), considerably less than the 4.8% measured for this particular MR-FORC, Fig. 5(b). The fact that two identical macroscopic magnetization values ($M/M_S=0$) can have quite different MR values further highlights that it is not the macroscopic magnetization value, but the microscopic magnetization domain structure and degree of total spin disorder, that is important in determining the MR for the [Co/Cu]$_8$ multilayer. In other words, a given macroscopic magnetization does not necessarily correlate with a unique MR value.

The difference in the reversal behavior of the [Co/Cu]$_8$ multilayer as compared to the PSV is that the constituent Co layers are on equal footings during reversal. Additionally, the Cu spacer thickness has been tuned to allow for a preferential anti-parallel coupling between adjacent Co layers.[7] As opposed to the conventional field-cycling along the major loop, during a MR-FORC measurement a large parameter space, spanned by both $H$ and $H_R$, is probed. In particular we find that those FORCs reversing near negative saturation can access higher spin-disordered states, assisted by the preferential antiferromagnetic interactions between Co layers, and result in larger MR values than possible from the standard major loop.

## V. CONCLUSIONS

We have investigated correlated magnetization reversal and magnetoresistance evolution in classical spintronic systems. Using a combined M and MR-FORC analysis we are able to access different spin-disordered states beyond what can be achieved under the conventional magnetic field cycling. The FORC analysis of the FePt/Cu/NiFe PSV showed a simple correlation between the magnetization and MR reversal processes. This is due to the large difference in switching fields between the FePt and NiFe and minimal interlayer interactions present. On the contrary, the behavior of the [Co/Cu]$_8$ multilayer was far more complex. Most notably, the irreversible switching processes, as evidenced in the M and MR-FORC-SFDs, showed no simple relationship, particularly for large negative reversal fields. This was also strikingly apparent in the MR-FORCs, which probed MR values larger than those found along the major loop. These differences arise because the switching fields of the individual Co layers are now more or less equal, allowing the finite interactions between neighboring Co layers, which favor anti-parallel alignment, to emerge during an MR-FORC measurement. Unlike the M-FORC measurements, which are only sensitive to the macroscopic changes in magnetization, the MR-FORC analysis is sensitive to the microscopic domain configurations and the net spin disorder, as those are most important in determining the resultant MR value. Our findings are also applicable to devices based on the much larger TMR effect,[44] or any given spintronic system, as a method to realize the maximum spin-disorder and the largest MR.

## ACKNOWLEDGMENTS

This work has been supported by the NSF (DMR-1008791 and ECCS-1232275), the Swedish Research Council (VR), the Swedish Institute (SI), the Swedish Foundation for Strategic Research (SSF), and the Knut and Alice Wallenberg Foundation.

**Figure Captions**

**Fig. 1.** (Color online) Major hysteresis loops of the magnetization (black solid squares) and magnetoresistance (red open circles) for the (a) FePt/Cu/NiFe pseudo spin valve and (b) [Co/Cu]$_8$ multilayer.

**Fig. 2.** (Color online) Families of (a) M-FORCs and (c) MR-FORCs for the FePt/Cu/NiFe PSV, whose starting points are represented by black dots. The insets highlight the low field reversal behavior. The corresponding FORC distributions are shown in (b) and (d), respectively. The dashed circle and oval highlight regions of the FORC distributions discussed in the text. (e) FORC-SFDs extracted from the M-FORC (black solid squares) and MR-FORC (red open circles) distributions.

**Fig. 3.** (Color online) Families of (a) M-FORCs and (c) MR-FORCs for the [Co/Cu]$_8$ multilayer, whose starting points are represented by black dots. The corresponding FORC distributions are shown in (b) and (d), respectively. (e) FORC-SFDs extracted from the M-FORC (black solid squares) and MR-FORC (red open circles) distributions.

**Fig. 4.** (Color online) MR major loops and selected MR-FORCs, starting from the indicated $H_R$ values, for the (a) FePt/Cu/NiFe PSV, (b) [Co/Cu]$_8$ multilayer, and (c) simulated Co/Cu/Co tri-layer. Simulated magnetization configurations in the top and bottom Co layers along the major loop at $H$= -23 Oe are shown in (c) inset, marked by a black dot on the major loop.

**Fig. 5.** (Color online) Selected M-FORCs (black solid squares) and MR-FORCs (red open circles) for the (a) simulated Co/Cu/Co tri-layer and (b) [Co/Cu]$_8$ multilayer. In (a) simulated domain configurations are shown for the top and bottom Co layers under different applied fields.

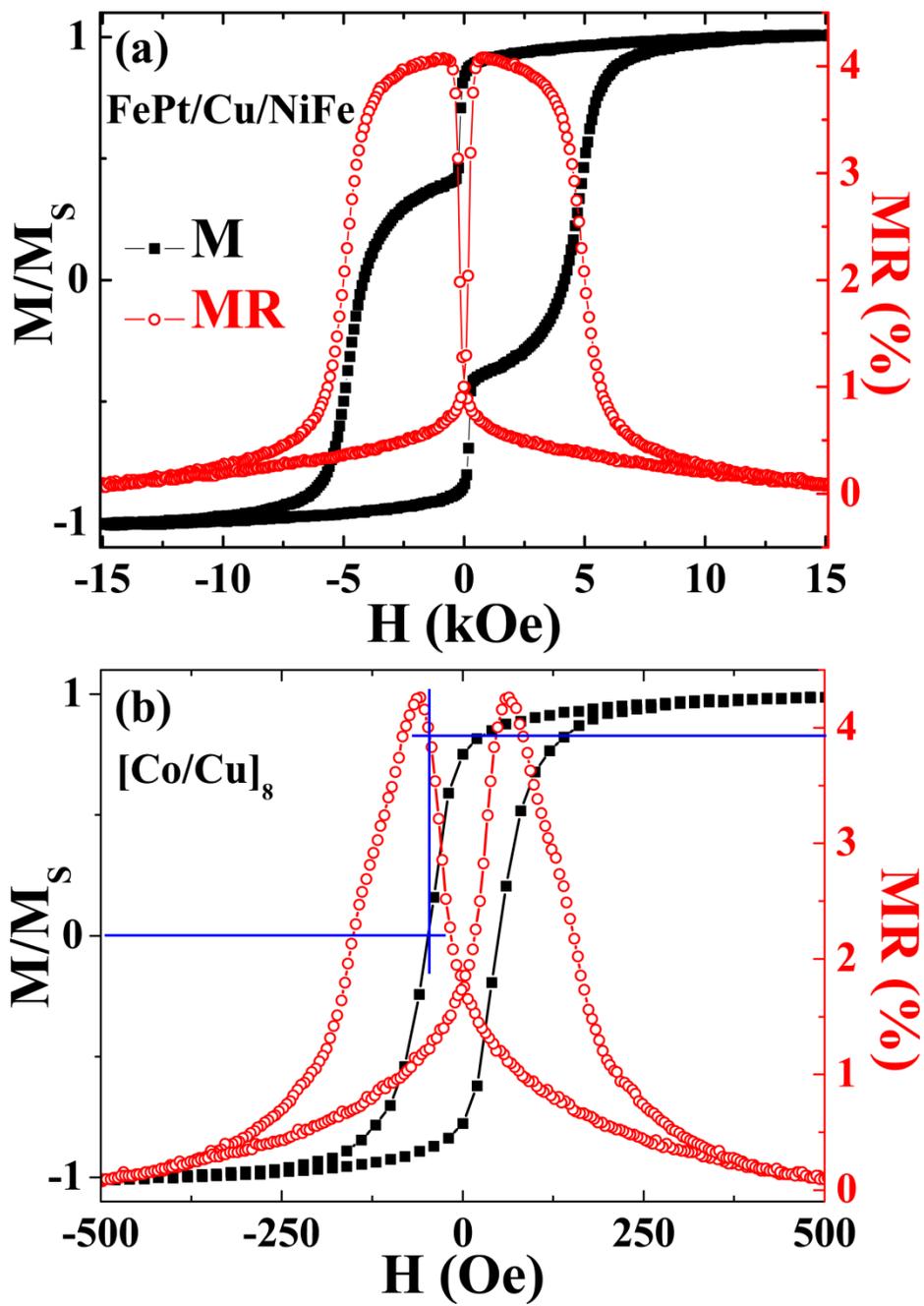

**Fig. 1.** Dumas, *et al.*

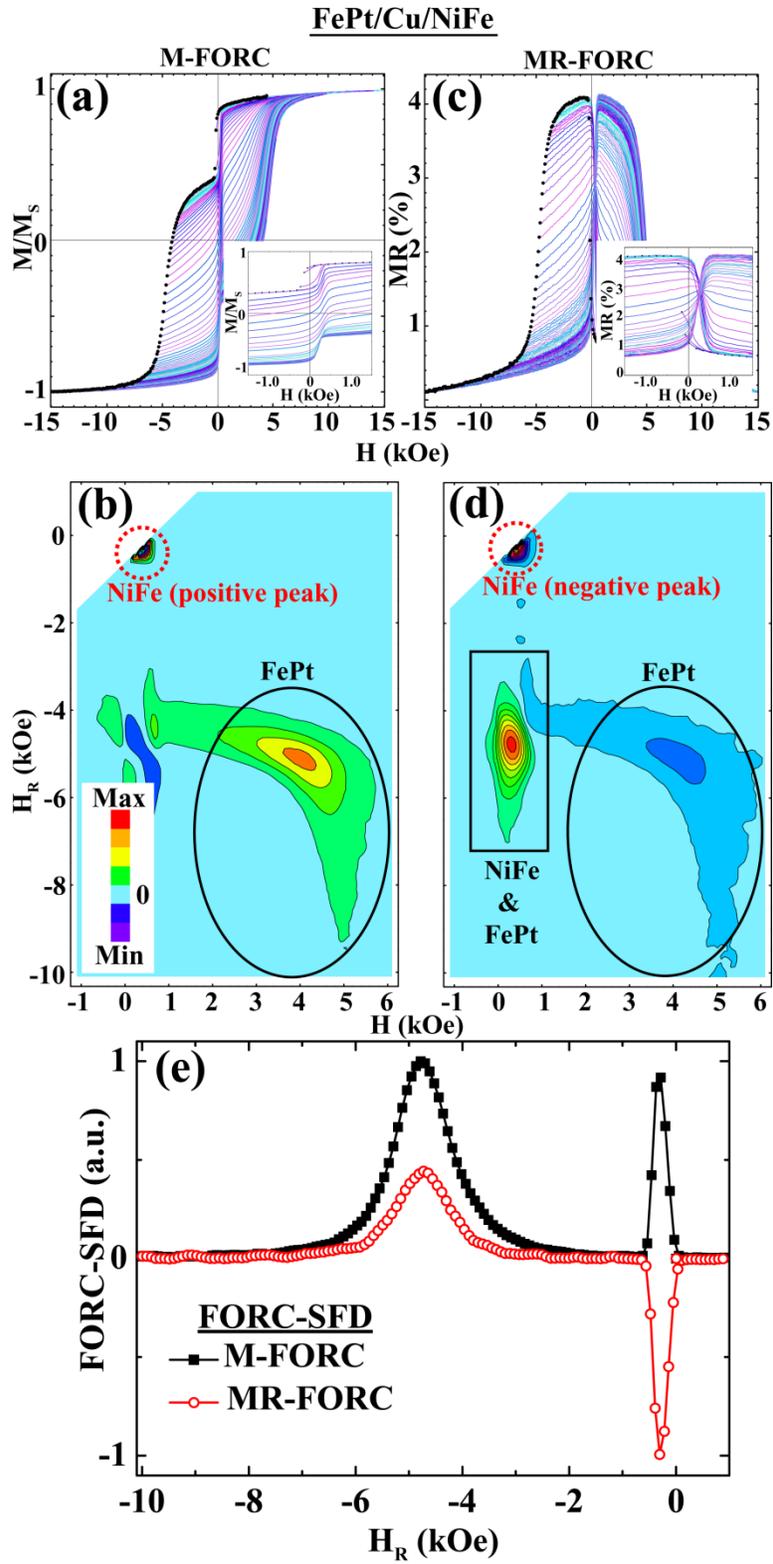

Fig. 2. Dumas, *et al.*

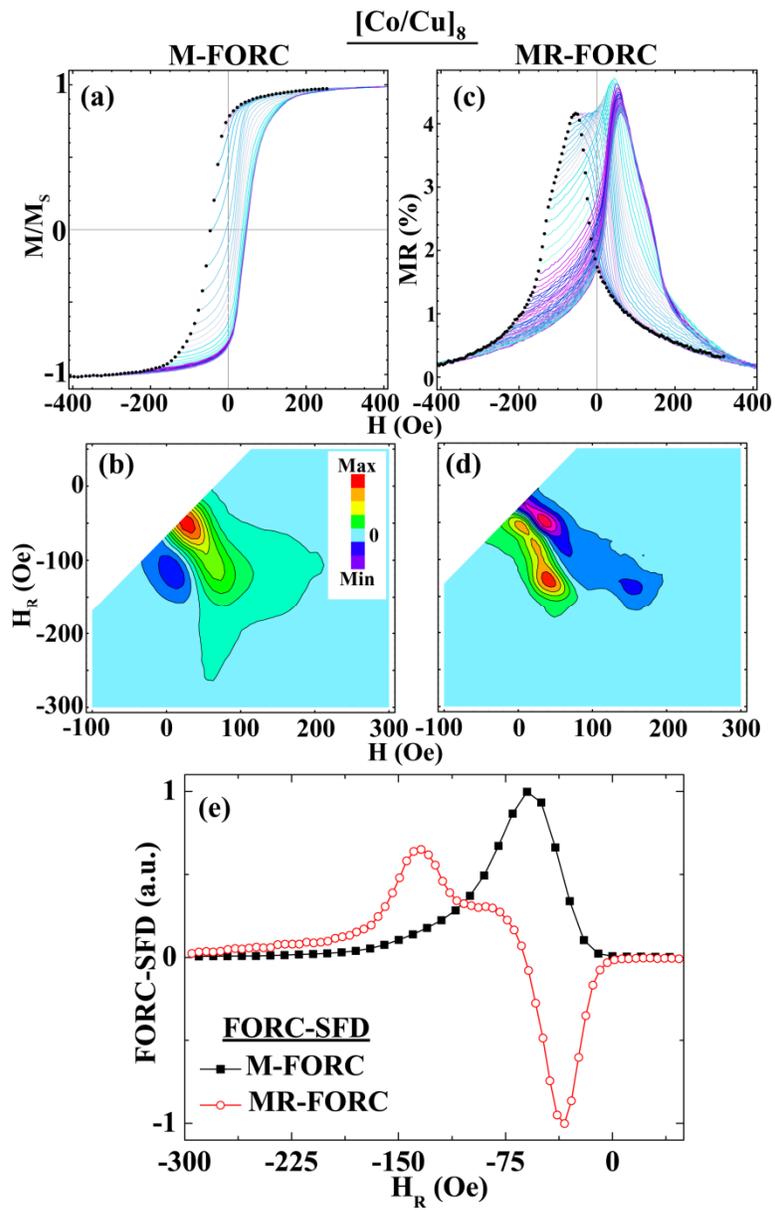

**Fig. 3.** Dumas, *et al.*

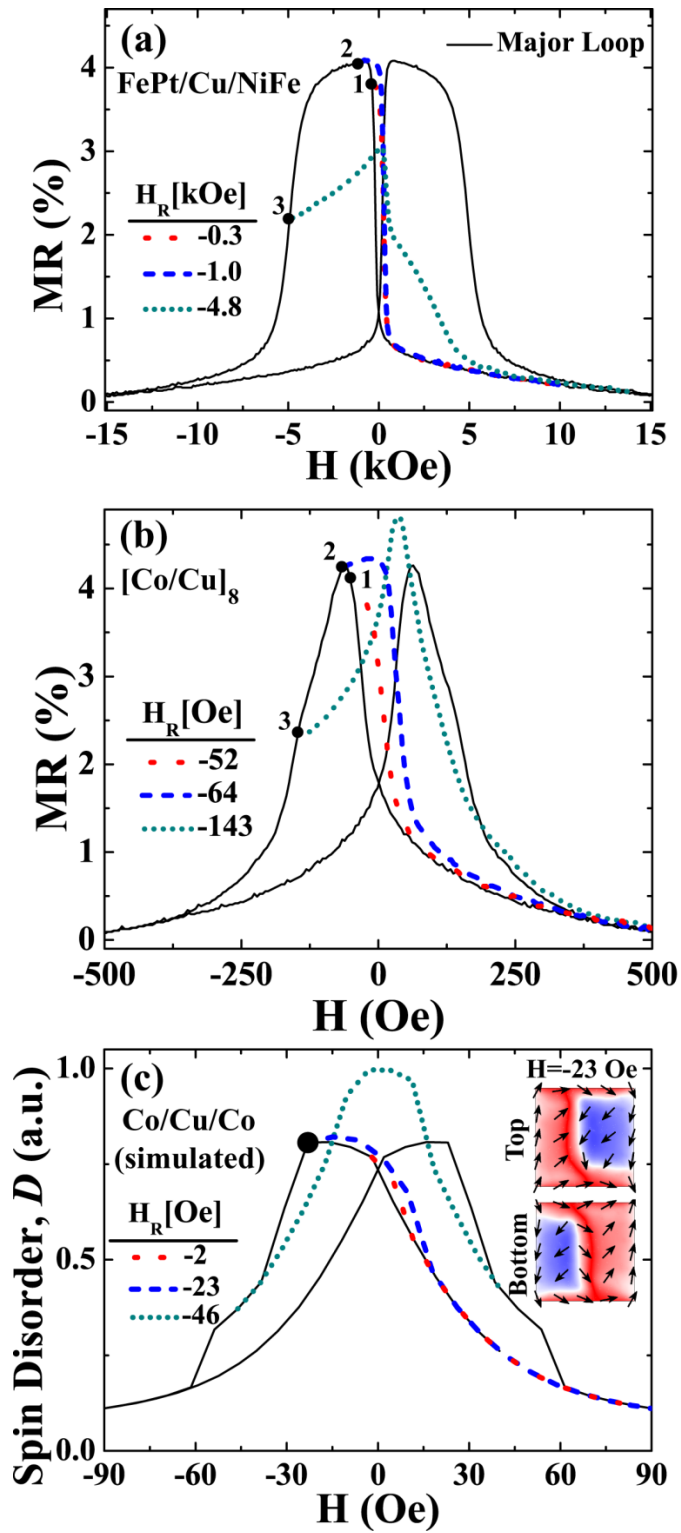

Fig. 4. Dumas, *et al.*

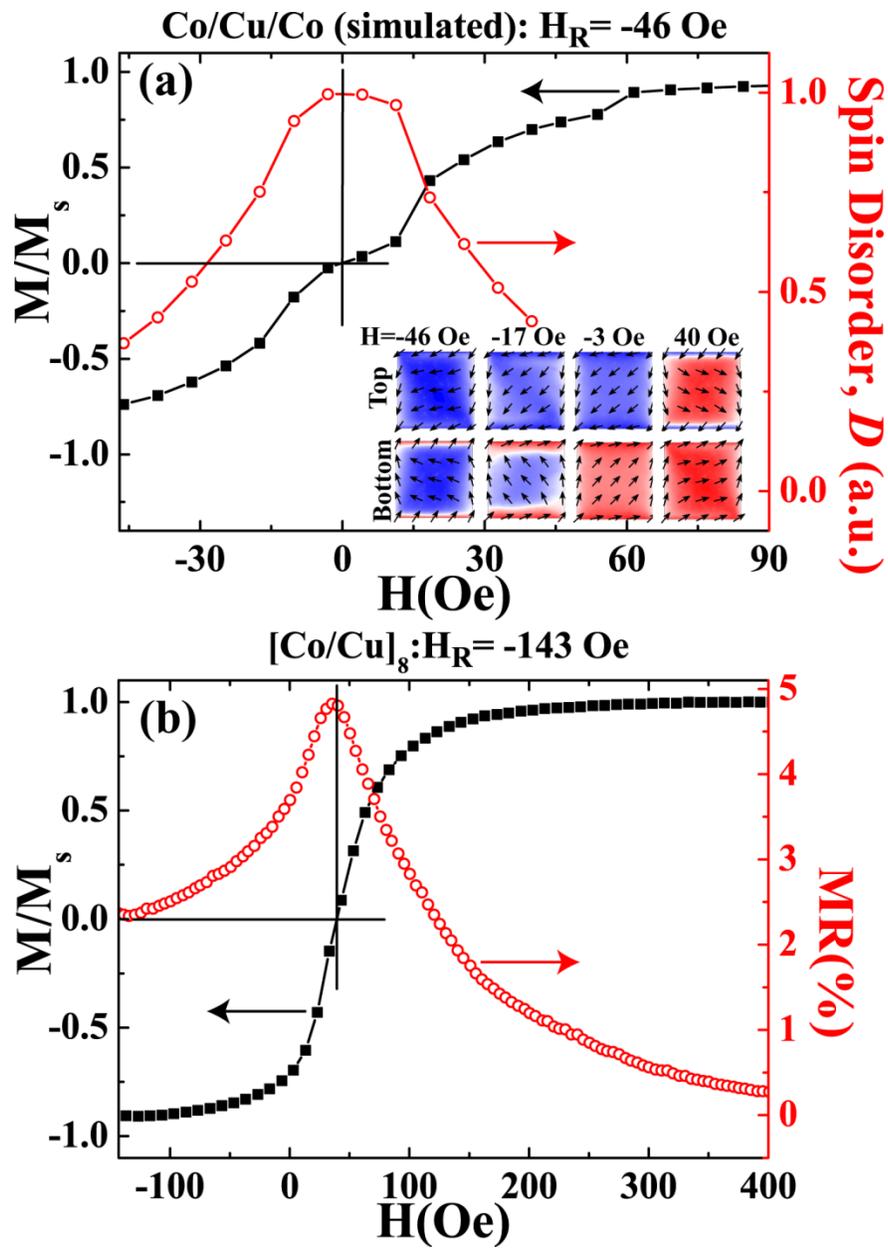

Fig. 5. Dumas, *et al.*